\documentclass[11pt,a4paper]{article}

\usepackage[margin=1in]{geometry}
\usepackage[T1]{fontenc}
\usepackage{microtype}
\usepackage{indentfirst}
\usepackage{amsmath,amssymb,mathtools,bm}

\usepackage{graphicx}
\usepackage{booktabs}
\usepackage{caption}
\usepackage[numbers,sort&compress]{natbib}

\usepackage[hidelinks]{hyperref}

\makeatletter
\renewcommand{\maketitle}{%
  \begingroup
  \centering

  \vspace*{-1.5em}

  \begin{minipage}{0.94\textwidth}
    \centering

    {\LARGE\bfseries
      \@title
      \par
    }

    \vspace{1.15em}

    {\large
      \@author
      \par
    }

    \vspace{0.85em}

    {\normalsize
      \@date
      \par
    }
  \end{minipage}

  \par
  \vspace{1.4em}
  \endgroup
}
\makeatother

\title{%
Low-Dimensional Reduction Theory for Populations of Phase Oscillators with a Gaussian Frequency Distribution%
}

\author{%
Kai Tokunaga\\[0.55em]
{\small
Department of Complexity Science and Engineering\\
The University of Tokyo, Kashiwa, Chiba 277-8561, Japan\\[0.35em]
\href{mailto:tokunaga.kai25a@c.k.u-tokyo.ac.jp}
{\nolinkurl{tokunaga.kai25a@c.k.u-tokyo.ac.jp}}%
}%
}

\date{27 August 2026}

\begin{document}

\maketitle

\begin{abstract}
Low-dimensional reduction theories such as the Ott--Antonsen ansatz have played a crucial role in the study of populations of coupled oscillators. Their application, however, has largely been restricted to systems with frequency distributions of rational-function form, such as the Cauchy distribution. For such distributions, the residue theorem allows the dynamics of the global order parameters to be closed in terms of a finite number of poles, thereby yielding a finite-dimensional system of ordinary differential equations. Rational frequency distributions, however, generally have heavy tails and only finitely many well-defined moments, and therefore may not always be realistic as frequency distributions. In this paper, we develop an approximate low-dimensional reduction theory based on perturbation theory for weakly heterogeneous populations of phase oscillators with a Gaussian frequency distribution. We construct the theory not only for first-harmonic coupling, for which the Ott–Antonsen ansatz applies, but also for populations of phase oscillators with multi-harmonic coupling. The effectiveness of the proposed low-dimensional reductions is demonstrated through both theoretical analysis and numerical simulations.
\end{abstract}

\section{Introduction}

Populations of coupled oscillators have long attracted attention because of their broad applications in biology, chemistry, engineering, physics, and other fields \cite{kuramoto1984chemical,winfree2001,Pikovsky_Rosenblum_Kurths_2001,strogatz2003sync}. One of the most prominent phenomena exhibited by populations of coupled oscillators is synchronization, in which a population of oscillators with heterogeneity spontaneously coordinates its oscillations. Phase reduction provides an important theoretical framework for studying the dynamics of populations of coupled oscillators, such as synchronization. Through phase reduction, the original complex system is greatly simplified to a population of phase oscillators described solely by their phase variables. In many cases, the heterogeneity among the oscillators is then represented by a distribution of their natural frequencies. Nevertheless, a population of phase oscillators with such a frequency distribution remains complex when the number of oscillators is large.

To analyze such populations, Ott and Antonsen developed an important theoretical framework. When the frequency distribution satisfies certain analyticity conditions, their theory reduces a population of sinusoidally coupled oscillators to a dynamical system on a low-dimensional invariant manifold \cite{Ott2008}. In particular, when the frequency distribution is of rational-function form, the residue theorem can be used to reduce the original system to a finite-dimensional system of ordinary differential equations. To date, low-dimensional reduction theories, such as the Ott–Antonsen ansatz, have played a central role in the study of oscillator populations and have undergone numerous developments in both exact \cite{Watanabe1993,Cestnik2022,Pietras2016,Martens2009,Skardal2018,tokunaga202608} and approximate \cite{Tyulkina2018,Vlasov_2016,tokunaga2026} forms.

However, many of these theories have been developed for systems with frequency distributions of rational-function form, for which a finite-dimensional reduction is possible. In general, rational-function distributions have only finitely many well-defined moments and possess heavy tails, and are therefore often unrealistic from the viewpoint of mathematical modeling. Moreover, the widely used Cauchy distribution has been reported to exhibit nonuniversal behavior \cite{Lafuerza2010,pietras2024}. For populations with more realistic frequency distributions, such as Gaussian distributions, approximations by rational functions have often been employed \cite{Campa2022,Klinshov2021}. However, this approach does not provide a specific error estimate for the accuracy with which the reduced equations approximate the dynamics of the original system.

n this paper, we develop a systematic low-dimensional reduction theory for oscillator populations whose frequencies follow a Gaussian distribution with a small standard deviation. The reduction is constructed from a perturbative hierarchy with explicit orders of approximation error. The theory is based on the Fourier–Hermite decomposition approach proposed by Chiba \cite{Chiba2013} and subsequently applied by León and Pazó \cite{Leon202204,Leon202206} to efficient numerical simulations. We show that the moments obtained from the Fourier–Hermite decomposition form a hierarchy ordered by successive powers of the standard deviation $\sigma$ of the Gaussian frequency distribution. The construction of the present theory is similar to that of the circular-cumulant approach \cite{Tyulkina2018} for systems subject to weak Gaussian noise; however, the variables forming the hierarchy of smallness are different, and the resulting reduced ODEs are also different. Furthermore, we develop the theory for systems with multi-harmonic coupling, which lies beyond the range of applicability of the OA ansatz. In the multi-harmonic case, the theory is constructed by combining the Fourier–Hermite approach with an approximate reduction based on the theory of orthogonal polynomials on the unit circle (OPUC) \cite{tokunaga2026}, which generalizes the OA ansatz.

\section{Problem Formulation}

In this paper, we consider a population of globally coupled phase oscillators with harmonics up to the $L$th harmonic, subject to independent normalized Cauchy noises of intensity $\gamma$, as follows:
\begin{equation}
\dot{\theta}_i = \omega_i + \sum_{l=1}^{L}[h_l(t)e^{il{\theta}_i}+\overline{h_l(t)}e^{-il{\theta}_i}] + \gamma \xi_i .
\end{equation}
The natural frequencies $\omega_i$ are assumed to be drawn from a Gaussian frequency distribution with mean $\omega_0$ and standard deviation $\sigma$.
We consider the thermodynamic limit of an infinite population. In this limit, the local phase density function $\rho(\theta,\omega,t)$ for each natural frequency $\omega$ obeys the following continuity equation:
\begin{equation}
\frac{\partial \rho(\theta,\omega, t)}{\partial t} + \frac{\partial}{\partial \theta} (h(\theta,\omega,t)\rho(\theta,\omega, t) ) = \gamma \left|\frac{\partial}{\partial \theta}\right|\rho(\theta,\omega, t).
\end{equation}
Here, the term on the right-hand side of Eq.~(2) represents the effect of Cauchy noise, and $\left|\frac{\partial}{\partial \theta}\right|$ is an operator that acts on each term in the Fourier expansion of a periodic function according to $\left|\frac{\partial}{\partial \theta}\right|e^{i n\theta}=-\left|n\right|e^{i n\theta}$.
We also define $h(\theta,\omega,t)=\omega + \sum_{l=1}^{L}[h_l(t)e^{il{\theta}_i}+\overline{h_l(t)}e^{-il{\theta}_i}]$.
The local moments determined from $\rho(\theta,\omega,t)$ for each $\omega$ are defined as $z_n(\omega,t) = \int_0^{2\pi} e^{-in\theta} \rho(\theta,\omega,t)\, d\theta$.
The global phase density of the entire population is given by $\rho(\theta,t) = \int_{-\infty}^{\infty} \rho(\theta,\omega,t)g(\omega)\, d\omega$, and its moments are given by $Z_n(t) = \int_{-\infty}^{\infty} z_n(\omega,t)g(\omega)\, d\omega$.
The phase density function $\rho(\theta,\omega,t)$ can be expanded in terms of the moments as
$
\rho(\theta,\omega,t)= \frac{1}{2\pi}\left\{1+\sum_{n=1}^{\infty} \bigl[z_n(\omega,t) e^{in\theta}+\overline{z_n(\omega,t)}e^{-in\theta}\bigr]\right\}.
$
Substituting this expansion into the continuity equation yields the following infinite set of ODEs for the moments:
\begin{equation}
 \dot{z}_n(\omega,t) = (-in\omega-\gamma|n|) z_n(\omega,t) - in \sum_{l=1}^{L}\bigl[h_l(t) z_{n-l}(\omega,t) +  \overline{h_l(t)} z_{n+l}(\omega,t)\bigr] \quad(n=-\infty,\dots,\infty).
\end{equation}

Alternatively, rather than expanding the phase density in terms of moments defined separately for each $\omega$, one can use a Fourier--Hermite expansion whose coefficients are independent of $\omega$ \cite{Chiba2013,Leon202204,Leon202206}. Let the normalized Hermite polynomials be $h_m(\omega)=\frac{\operatorname{He}_m(\frac{\omega-\omega_0}{\sigma})}{\sqrt{m!}}$, and define the moments of the Fourier--Hermite expansion by $P_n^{m}(t) = \int_{-\infty}^{\infty}\int_0^{2\pi} e^{-in\theta}h_m(\omega)g(\omega) \rho(\theta,\omega, t)\, d\theta d\omega$. The phase density can be expanded  as $
\rho(\theta,\omega, t)= \frac{1}{2\pi}\left\{1+\sum_{n=1}^{\infty}\sum_{m=0}^{\infty} \bigl[P_n^{m}(t) e^{in\theta}+\overline{P_n^{m}(t)}e^{-in\theta}\bigr]h_m(\omega)\right\}
$. Substituting this expansion into the continuity equation yields the following infinite set of ODEs for the moments:
\begin{equation}
\begin{split}
 \dot{P}_n^{m}(t) =&-in\sigma(\sqrt{m}P_n^{m-1}(t)+\sqrt{m+1}P_n^{m+1}(t)) -(in\omega_0+\gamma|n|) P_n^{m}(t) \\
 &- in \sum_{l=1}^{L}\bigl[h_l(t) P_{n-l}^{m}(t) +  \overline{h_l(t)} P_{n+l}^{m}(t)\bigr]\quad(n=-\infty,\dots,\infty,m=0,\dots,\infty).
 \end{split}
\end{equation}
Here, when $\sigma\ll1$, if $|P_n^{m}(t)|$ is assumed to decrease monotonically as $m$ increases, the term $in\sigma\sqrt{m}P_n^{m-1}(t)$ is found to be the dominant driving term in the ODE. It therefore follows that the moments form a hierarchy of smallness given by $P_n^{m}(t)=O(\sigma^m)$. Throughout this paper, we use $\omega$-independent initial conditions satisfying $P_n^m(0)=0\,(m\ge1)$. Comparing the expansions of $\rho(\theta,\omega, t)$, we obtain
\begin{equation}
z_n(\omega,t)=\sum_{m=0}^{\infty}P_n^{m}(t) h_m(\omega).
\end{equation}
Formally, this can be regarded as a perturbation expansion in small $\sigma$ around the moments $P_n^{0}(t)=Z_n(t)$ of the global phase density. The ODEs for $P_n^{m}\,(n=-\infty,\dots,\infty)$ describe the time evolution of the perturbation terms of order $O(\sigma^m)$. Since $P_n^{m}(t)=O(\sigma^m)$, the system obtained by truncating the hierarchy as $P_n^{m}=0\,(m\ge M+1)$ approximates the original system with an accuracy of $O(\sigma^{M+1})$. However, infinitely many terms still remain in the $n$ direction, and hence this does not yet constitute a finite-dimensional reduction. In this paper, we show that, as far as the long-time behavior is concerned, the hierarchy can be closed exactly using only $n=1$ through the OA ansatz when $L=1$, whereas for $L>1$ it can be approximated by a closure at $n=N\ge L$ using the OPUC-based method \cite{tokunaga2026}. We thereby demonstrate that, for a Gaussian frequency distribution with a small standard deviation, a perturbative low-dimensional reduction to a finite-dimensional system of ODEs is possible.
\section{Sinusoidal Coupling}
We consider the case of sinusoidal coupling described as follows:
\begin{equation}
 \dot{z}_n(\omega,t) = (-in\omega-\gamma|n|) z_n(\omega,t) - in \bigl[h_1(t) z_{n-1}(\omega,t) +  \overline{h_1(t)} z_{n+1}(\omega,t)\bigr].
\end{equation}
Eq.(6) has an invariant manifold (OA manifold) defined by $z_n(\omega,t)=z_1(\omega,t)^n$. This invariant manifold is infinite-dimensional in the function space to which $\rho(\theta,\omega,t)$ belongs. Based on results for systems equivalent to Eq.~(6) \cite{Cestnik2022}, the OA manifold of the present system is expected to be globally stable when $\gamma>0$. Even when $\gamma=0$, under appropriate analyticity conditions on the initial condition, asymptotic stability has been established for the OA manifold after averaging over the natural frequency $\omega$ \cite{Ott2009,Ott2011}.

We now express the ODEs on the OA manifold in terms of $P_n^m(t)$. Since $z_{n+1}(\omega,t)=z_n(\omega,t)z_1(\omega,t)$, substituting Eq.~(5), multiplying by $g(\omega)h_m(\omega)$, and integrating yield
\begin{equation}
 P_{n+1}^m = \sum_{r,s=0}^{\infty}C_{mrs}P_n^rP_1^s.
\end{equation}
\begin{equation}
 C_{mrs}=   \begin{cases}
     \frac{\sqrt{r!s!m!}}{q!(r-q)!(s-q)!}& \text{if $m=r+s-2q$,} \\
    0                 & \text{otherwise.}
  \end{cases}
\end{equation}
To construct an approximation with accuracy $O(\sigma^{M+1})$, the following relation for $n=2$ is sufficient for the closure:
\begin{equation}
 P_{2}^m = \sum_{r,s=0}^{M}C_{mrs}P_1^rP_1^s.
\end{equation}
Using this relation, the ODEs can be written as
\begin{equation}
\begin{split}
 \dot{P}_1^{m}(t) =&-i\sigma(\sqrt{m}P_1^{m-1}(t)+\sqrt{m+1}P_1^{m+1}(t))\\& -(i\omega+\gamma) P_1^{m}(t) - i\bigl[\delta_{m0}h_1(t) +  \overline{h_1(t)} P_{2}^{m}(t)\bigr] \quad(m=0,...,M).
 \end{split}
\end{equation}
For $M=0$, this gives an ODE on the two-dimensional manifold defined by $z_n(\omega,t)=Z_1(t)^n$. This manifold is simply the OA manifold at $\sigma=0$. Therefore, this reduction approximates the dynamical system on the infinite-dimensional OA manifold by perturbation theory in small $\sigma$ around the two-dimensional OA manifold at $\sigma=0$. The finite-dimensional system obtained in this way approximates the original system with accuracy $O(\sigma^{M+1})$. However, for $P_n^m\,(0\le m\le M)$, the approximation has a higher accuracy. This is because $C_{mrs}$ is nonzero only when $|r-s|\le m\le r+s$ and $m-r-s$ is even. It therefore follows that the largest term neglected in
$
 P_{2}^m = \sum_{r,s=0}^{M}C_{mrs}P_1^rP_1^s
$
is of order $O(\sigma^{2(M+1)-m})$. Furthermore, in the term $\sigma(\sqrt{m}P_1^{m-1}(t)+\sqrt{m+1}P_1^{m+1}(t))$, the first moment affected by the truncation is $P_1^{M+1}=O(\sigma^{M+1})$. Each decrease in $m$ introduces an additional factor of $\sigma$, and hence the error arising from this term is also $O(\sigma^{2(M+1)-m})$. In many cases, the quantity of interest is the global phase density. Since the global phase density is determined by $P_k^0\,(m=-\infty,\dots,\infty)$, its error is found to be $O(\sigma^{2(M+1)})$.
\begin{figure}[t]
    \centering
    \includegraphics[width=1.0\linewidth]{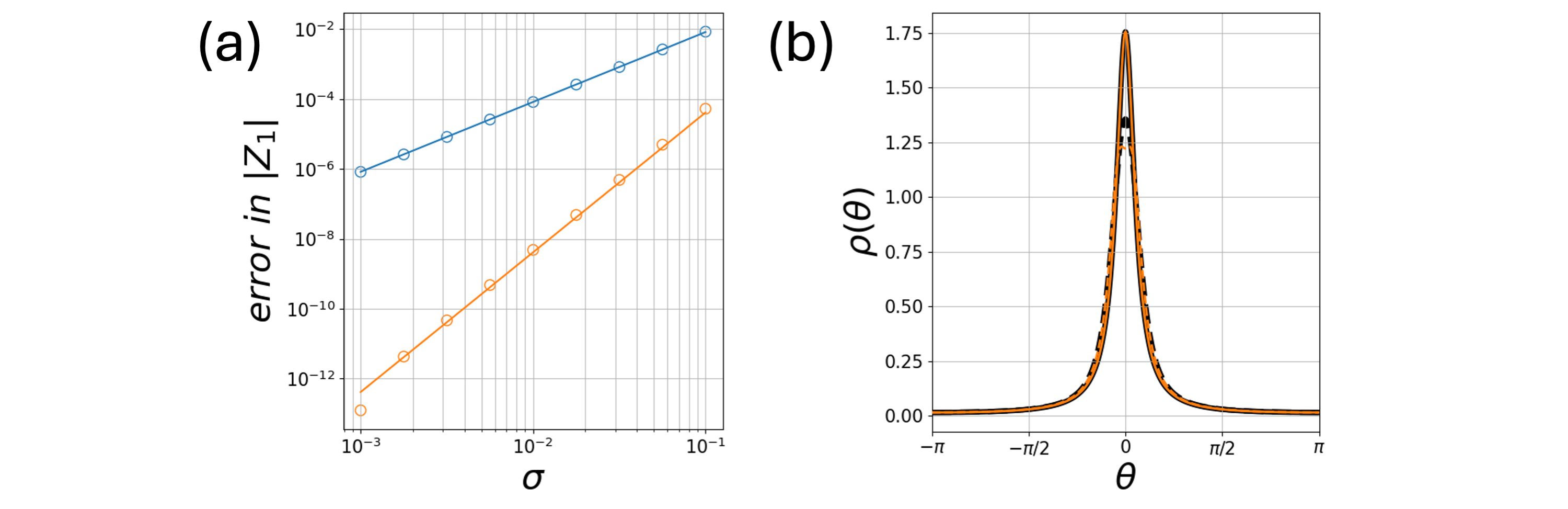}
    \caption{(a) The errors of the low-dimensional reduced systems relative to the original system are plotted for the stationary solution of the Kuramoto model. Blue and orange indicate $M=0$ and $M=1$, respectively. In the log--log plot, a straight line with slope 2 proportional to $\sigma^2$ (blue) and a straight line with slope 4 proportional to $\sigma^4$ (orange) are drawn to fit the data points for $M=0$ and $M=1$, respectively. (b) The phase density of the original system (black) is compared with that of the reduced system with $M=1$ (orange). The solid and dashed lines correspond to $\sigma=0.001$ and $\sigma=0.1$, respectively.}
\end{figure}

Next, we verify the validity of the reduction through numerical simulations. We examine the accuracy of the following system with $M=1$.
\begin{equation}
\begin{split}
&\dot{P}_1^{0}(t) =-i\sigma P_1^{1}(t) -(i\omega_0+\gamma) P_1^{0}(t) - i\bigl[h_1(t) +  \overline{h_1(t)} (P_{1}^{0}(t)^2+P_{1}^{1}(t)^2)\bigr] \\
&\dot{P}_1^{1}(t) =-i\sigma P_1^{0}(t) -(i\omega_0+\gamma) P_1^{1}(t) - 2i\overline{h_1(t)} P_{1}^{0}(t)P_{1}^{1}(t)
\end{split}
\end{equation}
Here, for the Kuramoto model with mean natural frequency $\omega_0=0$, given by $h_1(t)=-\frac{\kappa_1}{2i}Z_1(t)$, the synchronized state is a stationary solution. For this stationary solution, we simulate a reference system to obtain $Z_1(t)=P_1^{0}(t)$ and examine the errors of the reduced systems obtained by approximating the ODEs on the OA manifold with $M=0$ and $M=1$. Throughout this paper, we use a Fourier--Hermite system with $N=300$ and $M=30$ as the reference system, and linear extrapolation is used for the truncation in the Hermite direction, following the method adopted in previous work \cite{Leon202206}. Figure~1(a) shows the results. It is confirmed that the error estimates $O(\sigma^2)$ for $M=0$ and $O(\sigma^4)$ for $M=1$ hold. Figure~1(b) shows the corresponding phase densities. For $\sigma=0.001$ and $0.1$, the results obtained from the reference system and the system with $M=1$ are compared.

The reduction developed here for systems with independent Cauchy noise and a Gaussian frequency distribution with a small standard deviation has a theoretical structure closely analogous to that of the circular-cumulant approach for sinusoidally coupled phase-oscillator populations with a Cauchy frequency distribution and weak Gaussian noise \cite{Tyulkina2018}, and can be regarded as its counterpart. However, the variables that form the hierarchy of smallness are not circular cumulants but the moments of the Fourier--Hermite expansion, and the resulting reduced equations also take a different form. For sinusoidally coupled phase-oscillator populations, it is known that a system subject to independent Cauchy noise is equivalent to one with a Cauchy frequency distribution \cite{Cestnik2022}. However, this equivalence no longer holds between a system with independent Cauchy noise and a Gaussian frequency distribution and one with independent Gaussian noise and a Cauchy frequency distribution. Indeed, for small $\sigma$, the squared amplitudes of the synchronized solutions of the Kuramoto model, $|Z_1|_{\omega}^2$ for independent Cauchy noise of intensity $\gamma$ and a Gaussian frequency distribution with mean zero and variance $\sigma^2$, and $|Z_1|_{n}^2$ for independent Gaussian noise of intensity $\sigma^2$ and a Cauchy frequency distribution centered at zero with width $\gamma$, are respectively given by
\begin{equation}
 |Z_1|_{\omega}^2 =1-\frac{2\gamma}{K}-\frac{\sigma^2}{(K-\gamma)^2}+O(\sigma^4),
\end{equation}
\begin{equation}
 |Z_1|_{n}^2 =1-\frac{2\gamma}{K}-\frac{\sigma^2}{K-\gamma}+O(\sigma^4).
\end{equation}
Thus, the two expressions already differ at order $O(\sigma^2)$.

\section{Multi-Harmonic Coupling}
We consider the case of multi-harmonic coupling with $L\ge1$. We assume $\gamma>0$ and Kuramoto--Sakaguchi-type coupling. Namely, denoting the coupling strength and phase shift by $\kappa_l$ and $\tau_l$, respectively, we write $h_l(t)=-\frac{\kappa_l e^{i\tau_l}}{2i}Z_l(t)$. First, we consider a uniformly rotating solution in which the global phase density $\rho(\theta,t)$ rotates with frequency $\Omega$. In a reference frame rotating with frequency $\Omega$, the global phase density can be regarded as a stationary solution, $\rho(\theta,t)^{(st)}=\rho(\theta-\Omega t)^{(st)}$. When $\rho(\theta-\Omega t,\omega,t)^{(st)}$ is decomposed using the Fourier--Hermite expansion, the time derivative vanishes from the ODE for $m=0$ because the global phase density is stationary, yielding
\begin{equation}
 0=-in\sigma P_n^{1}(t) -(in(\omega_0-\Omega)+\gamma|n|) P_n^{0(st)} - in \sum_{l=1}^{L}\bigl[h_l^{(st)} P_{n-l}^{0(st)} +  \overline{h_l^{(st)}} P_{n+l}^{0(st)}\bigr].
\end{equation}
It follows from this equation that $P_n^{1}(t)=P_n^{1(st)}$. Next, the ODE for $m=1$ gives $P_n^{2}(t)=P_n^{2(st)}$. Repeating this procedure successively, we find inductively that $P_n^{m}(t)=P_n^{m(st)}$ holds for all $m$. From the relation between $P_n^m(t)$ and $z_n(\omega,t)$, it then follows that $z_n(\omega,t)=z_n(\omega)^{(st)}$ for all $n$. Thus, $z_n(\omega,t)$ is stationary in the rotating frame and undergoes uniform rotation with frequency $\Omega$ in the original frame. We then obtain the following recurrence relation:
\begin{equation}
  (\omega-\Omega - i\gamma) z^{(st)}_n(\omega)
  + \sum_{l=1}^{L}\bigl[\,h^{(st)}_l z^{(st)}_{n-l}(\omega) + \overline{h^{(st)}_l} z^{(st)}_{n+l}(\omega)\bigr] = 0 \quad(n\ge 1).
\end{equation}
This recurrence relation is equivalent to the one that appears in the low-dimensional reduction for multi-harmonic coupling \cite{Tonjes2020,tokunaga2026}, with the only difference being that the recurrence relation is defined separately for each $\omega$. Therefore, the Verblunsky coefficients for each $\omega$ satisfy $\alpha_n(\omega)=0\,(n\ge L)$, and $\rho(\theta,\omega,t)^{(st)}$ lies on the manifold represented by Bernstein--Szeg\H{o} measures (BS manifold) of degree $L$ . 

Next, we consider nonequilibrium solutions. Suppose that the global phase density $\rho(\theta,t)$ follows a nonequilibrium solution and varies sufficiently slowly in time in a frame rotating at the mean frequency $\Omega$ of the nonequilibrium solution. Introducing the slow time $\tau=\varepsilon t$, we assume that $\rho^{(qst)}(\theta,t)=\rho^{(qst)}(\theta-\Omega t,\tau)$. Here, since $P_n^{0}$ is a moment of the global phase density, $P_n^{0\,(qst)}(t)=P_n^{0\,(qst)}(\tau)$. By successively considering the ODEs for $P_n^m(t)$, it follows inductively that $P_n^{m\,(qst)}(t)=P_n^{m\,(qst)}(\tau)$ holds for all $m$. It can then be shown that $z_n^{(qst)}(\omega,t)=z_n^{(qst)}(\omega,\tau)$. Expanding $z_n^{(qst)}(\omega,\tau)=z_n^{(qst)\,(0)}(\omega,\tau)+\varepsilon z_n^{(qst)\,(1)}(\omega,\tau)+O(\varepsilon^2)$, we obtain the following recurrence relations:
\begin{equation}
 O(1):\, (\omega-\Omega - i\gamma) z_n^{(qst)\,(0)}(\omega,\tau)
  + \sum_{l=1}^{L}\bigl[\,h^{(st)}_l z^{(qst)\,(0)}_{n-l}(\omega,\tau) + \overline{h^{(st)}_l} z^{(qst)\,(0)}_{n+l}(\omega,\tau)\bigr] = 0 \quad(n\ge 1)
\end{equation}
\begin{equation}
O(\varepsilon):\, i\sum_{m=1}^{L} A_m\lambda_m^{n}+i\sum_{m=1}^{L} \frac{dC_m}{d\tau}\frac{\lambda_m^{n}}{n} =(\omega-\Omega - i\gamma) u_n
  + \sum_{l=1}^{L}\bigl[\,h^{(qst)}_l u_{n-l}+ \overline{h^{(qst)}_l}u_{n+l}\bigr] \quad  (n \ge 1)
\end{equation}
Here, $\lambda_m$ are the roots inside the unit disk of the linear recurrence relation at $O(1)$, and $A_m$ collectively represents the coefficients of the inhomogeneous terms proportional to $\lambda_m^{n}$. These recurrence relations are equivalent to those appearing in the OPUC reduction for multi-harmonically coupled phase-oscillator populations \cite{tokunaga2026}. Therefore, defining $\Lambda_{\max}(\omega)=\max_{1\le m\le L,\ t>T}|\lambda_m(\omega,t)|$, the Verblunsky coefficients for each $\omega$ satisfy $\alpha_n(\omega,t)=O(\frac{\varepsilon\Lambda_{\max}^n(\omega)}{n^2})\,(n\ge L)$. Note, however, that the accuracy deteriorates when the roots inside the unit disk approach one another. In deriving the quasistationary approximation, we assumed that the nonequilibrium solution varies slowly in time, but this assumption is not necessarily required. More generally, it is expected to be sufficient that the temporal variation of the nonequilibrium solution be sufficiently slow compared with relaxation toward the leading-order quasistationary manifold. The derivation of analogous results under such a more general assumption will be reported elsewhere.

Therefore, defining $\overline{\Lambda}_{\max}=\max_{\omega \in \mathbb{R}}|\Lambda_m(\omega)|$, we find that the system obtained by setting $\alpha_n(\omega,t)=0\,(n\ge N)$ for $N\ge L$ exactly reproduces the uniformly rotating solutions of the original system and approximates nonequilibrium solutions with an accuracy of $O(\frac{\varepsilon\overline{\Lambda}_{\max}^N}{N^2})$. Furthermore, truncating the hierarchy in the $m$ direction by setting $P_n^m(t)=0\,(m\ge M+1)$ yields a finite-dimensional system of dimension $2N(M+1)$, which approximates the original system with an accuracy of $O(\sigma^{(M+1)}+\frac{\varepsilon\overline{\Lambda}_{\max}^N}{N^2})$. In particular, uniformly rotating solutions are approximated with an accuracy of $O(\sigma^{(M+1)})$. When $M=0$, this becomes a reduction onto the $2N$-dimensional BS manifold at $\sigma=0$. Therefore, the reduction to this $2N(M+1)$-dimensional finite-dimensional system perturbatively approximates the dynamical system on the infinite-dimensional BS manifold in small $\sigma$, using the $2N$-dimensional BS manifold at $\sigma=0$ as the reference.

\begin{figure}[t]
    \centering
    \includegraphics[width=1.0\linewidth]{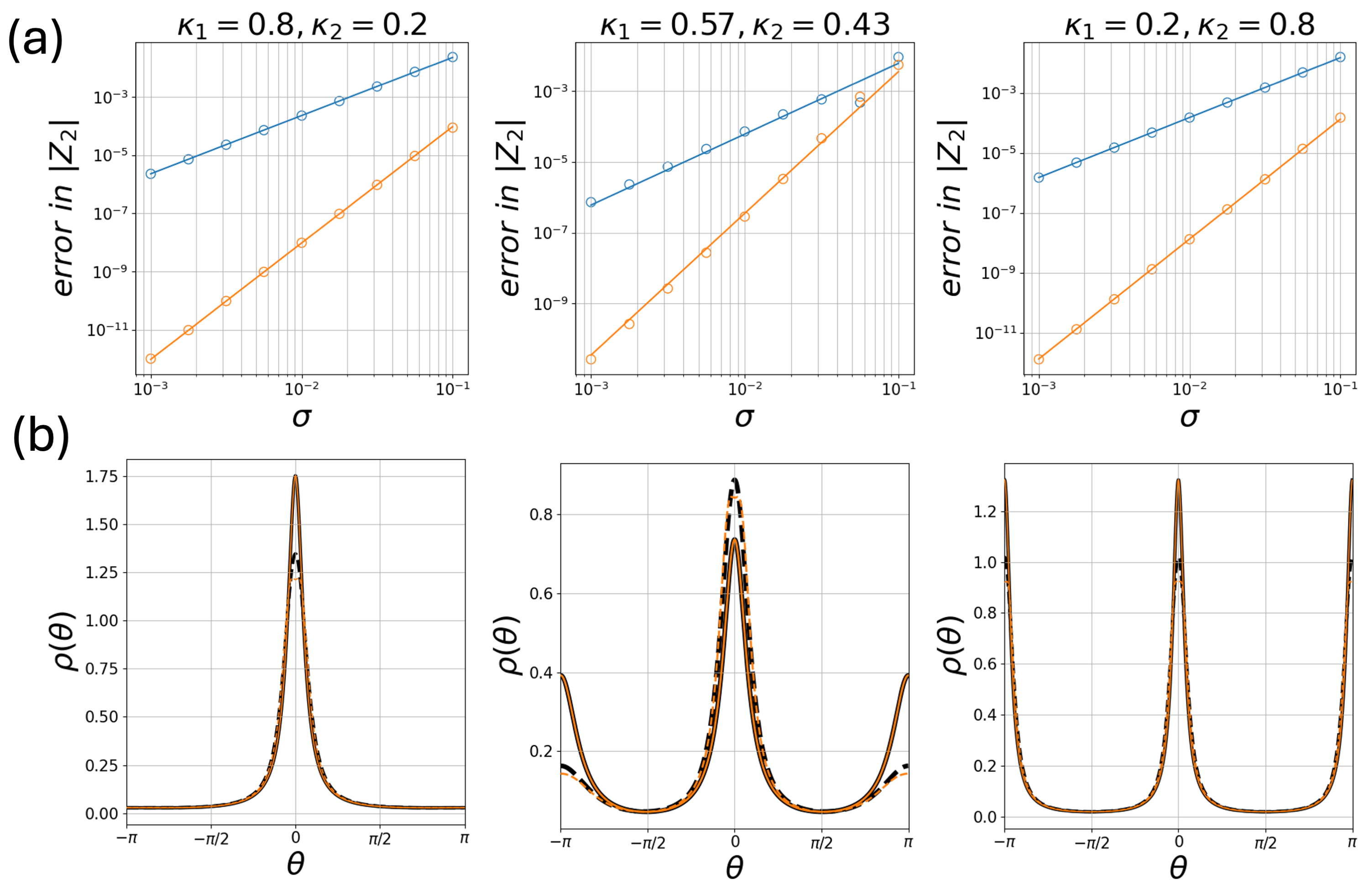}
    \caption{(a) The errors of the low-dimensional reduced systems relative to the original system are plotted for the stationary solutions of the model with $L=2$ and no phase shifts. Blue and orange indicate $M=0$ and $M=1$, respectively. In the log--log plot, a straight line with slope 2 proportional to $\sigma^2$ (blue) and a straight line with slope 4 proportional to $\sigma^4$ (orange) are drawn to fit the data points for $M=0$ and $M=1$, respectively. (b) The phase density of the original system (black) is compared with that of the reduced system with $M=1$ (orange). The solid and dashed lines correspond to $\sigma=0.001$ and $\sigma=0.1$, respectively.}
\end{figure}

Based on this $2N(M+1)$-dimensional reduction, we now explicitly construct a closure for the following ODEs:
\begin{equation}
\begin{split}
 \dot{P}_n^{m}(t) =&-in\sigma(\sqrt{m}P_n^{m-1}(t)+\sqrt{m+1}P_n^{m+1}(t)) -(in\omega_0+\gamma|n|) P_n^{m}(t) \\
 &- in \sum_{l=1}^{L}\bigl[h_l(t) P_{n-l}^{m}(t)(\omega,t) +  \overline{h_l(t)} P_{n+l}^{m}(t)\bigr] \quad(n=1,\dots,N,m=0,\dots,M).
 \end{split}
\end{equation}
For the system obtained by setting $\alpha_n(\omega,t)=0\,(n\ge N)$, let the reversed OPUC of degree $N$ be $\Phi_N^*(z,\omega,t)=1+\sum_{k=1}^{N}d_{k}(\omega,t) z^k$. The coefficients $d_k$ are then determined from the Toeplitz system
\begin{equation}
z_n(\omega,t)=-\sum_{k=1}^{N}d_{k}(\omega,t) z_{n-k}(\omega,t) \quad(1\le n\le N),
\end{equation}
after which the higher-order moments can be closed successively as follows:
\begin{equation}
z_n(\omega,t)=-\sum_{m=1}^{N}d_{m}(\omega,t) z_{n-m}(\omega,t) \quad(N+1\le n\le N+L).
\end{equation}
Here, $z_n(\omega,t)$ can be expanded in Hermite polynomials as $z_n(\omega,t)=\sum_{m=0}^{\infty}P_n^{m}(t) h_m(\omega)$, where the coefficients of $h_m(\omega)$ satisfy $P_n^{m}(t)=O(\sigma^m)$. Moreover, $d_m(\omega,t)$ is a function of $z_{-N}(\omega,t),\dots,z_N(\omega,t)$. Therefore, expanding $d_k(\omega,t)=\sum_{m=0}^{\infty}D_k^{m}(t)h_m(\omega)$, we find that $D^{m}(t)=O(\sigma^m)$. Since terms of order $O(\sigma^{M+1})$ are neglected in the present reduction, it is sufficient to set $z_n(\omega,t)=\sum_{m=0}^{M}P_n^{m}(t) h_m(\omega)$ and $d_{m}(\omega,t)=\sum_{m=0}^{M}D_k^{m}(t) h_m(\omega)$. Substituting these expansions into the Toeplitz system, multiplying by $h_m(\omega)g(\omega)$, and integrating yield
\begin{equation}
P_n^m(t)=-\sum_{k=1}^{N}\sum_{a=0}^{M}\sum_{b=0}^{M}C_{mab}D_{k}^a(t) P_{n-k}^b(t) \quad(1\le n\le N).
\end{equation}
Defining the matrices $\{\bm{L}_q\}_{m,a}=\sum_{b=0}^{M}C_{mab}P_{q}^b(t)$ and the vectors $\bm{D}_j=(D_j^0,\dots,D_j^M)$ and $\bm{P}_k=(P_k^0,\dots,P_k^M)$, we obtain the following block Toeplitz system:
\begin{equation}
 \begin{pmatrix}
\bm{L}_0   & \bm{L}_1     & \cdots & \bm{L}_{N-1} \\
\bm{L}_{-1}& \bm{L}_0     &        & \bm{L}_{N-2} \\
\vdots& \vdots  &        & \vdots \\
\bm{L}_{-N+1}& \bm{L}_{-N+2}& \cdots & \bm{L}_0
\end{pmatrix} 
\begin{pmatrix}
\bm{D}_N \\
\vdots\\
\bm{D}_1
\end{pmatrix}=-
\begin{pmatrix}
\bm{P}_N \\
\vdots\\
\bm{P}_{1}
\end{pmatrix}.
\end{equation}
Solving this system for $D_k^a$, the higher-order moments can be closed successively as follows:
\begin{equation}
P_n^m(t)=-\sum_{k=1}^{N}\sum_{a=0}^{M}\sum_{b=0}^{M}C_{mab}D_{k}^a(t) P_{n-k}^b(t) \quad(N+1\le n\le N+L).
\end{equation}
Considering the condition under which $C_{mab}$ is nonzero in this closure, we find, as in the case $L=1$, that the largest neglected term is of order $O(\sigma^{2(M+1)-m})$. Applying the same argument as for $L=1$ to the term $\sigma(\sqrt{m}P_1^{m-1}(t)+\sqrt{m+1}P_1^{m+1}(t))$, we conclude that the error in $P_n^m$ is $O(\sigma^{2(M+1)-m})$ for uniformly rotating solutions and $O(\sigma^{2(M+1)-m}+\frac{\varepsilon\overline{\Lambda}_{max}^N}{N^2})$ for nonequilibrium solutions.

\begin{figure}[t]
    \centering
    \includegraphics[width=1.0\linewidth]{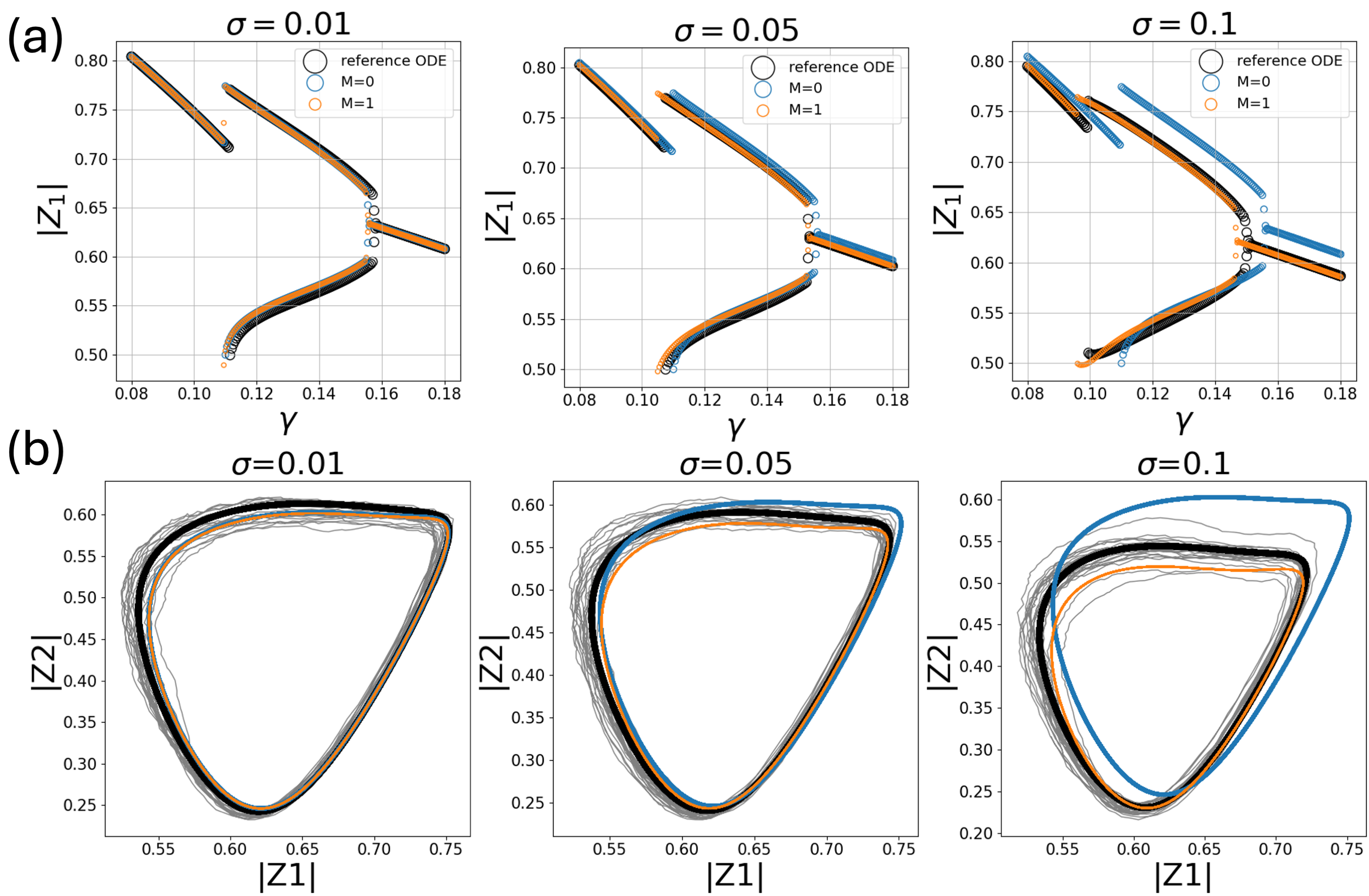}
    \caption{(a) In the parameter region where breathing behavior is observed for $L=3$, the maximum and minimum values of $|Z_1|$, obtained by decreasing $\gamma$ for different values of $\sigma$, are shown for the reference system (black), the reduced system with $M=0$ (blue), and the reduced system with $M=1$ (orange). (b) The corresponding periodic orbits at $\gamma=0.12$ are plotted in the $|Z_1|$--$|Z_2|$ plane. In addition to the reference system (black), the reduced system with $M=0$ (blue), and the reduced system with $M=1$ (orange), the particle simulation using the Langevin equations for the phases is shown in gray.}
\end{figure}

Next, we verify the validity of the reduction through numerical simulations. We set $L=2$, the phase shifts to $\tau_1=\tau_2=0$, and the mean natural frequency to $\omega_0=0$. In this case, synchronization and clustering appear as stationary solutions. For these stationary solutions, we simulate the reference system to obtain the actual value of $Z_2(t)=P_2^{0}(t)$ and evaluate, for each $\sigma$, the errors relative to the values obtained by simulating the approximate systems with $N=2$ and $M=0,1$. Figure~2(a) shows the results for different values of $(\kappa_1,\kappa_2)$ satisfying $\kappa_1+\kappa_2=1$. It is confirmed that the error estimates $O(\sigma^2)$ for $M=0$ and $O(\sigma^4)$ for $M=1$ hold. Figure~2(b) shows the phase densities corresponding to the respective values of $(\kappa_1,\kappa_2)$ in Fig.~2(a). For $\sigma=0.001$ and $0.1$, the results obtained from the reference system and the system with $M=1$ are compared.

We next perform numerical simulations for nonequilibrium solutions. For $\kappa_1=1.0$, $\kappa_2=0.7$, $\kappa_3=0.5$, $\tau_1=-0.2$, $\tau_2=-1.0$, and $\tau_3=-0.9$, breathing behavior emerges through a Hopf bifurcation as $\gamma$ is decreased and disappears through a SNIC bifurcation. The maximum and minimum values of $|Z_1|$ are compared between the reference system and the low-dimensional reduced systems with $N=5$ and $M=0,1$. The region in which the maximum and minimum values differ corresponds to periodic orbits. Figure~3(a) shows the results for $\sigma=0.01$, $0.05$, and $0.1$. Since the effect of $\sigma$ is not included when $M=0$, the results have the same form for all three values of $\sigma$. As $\sigma$ increases, the system with $M=1$ provides a better approximation than that with $M=0$. In Fig.~3(b), the periodic orbit associated with the breathing behavior is compared among the large-scale reference system, the particle simulation, and the low-dimensional reduced systems with $N=2L-1$ and $M=0,1$.

\section{Conclusion}
In this paper, we developed an approximate low-dimensional reduction theory based on perturbation theory for populations of phase oscillators with a Gaussian frequency distribution characterized by a small standard deviation. For sinusoidal coupling, the theory approximates the dynamical system on the infinite-dimensional Ott–Antonsen (OA) manifold through a perturbative expansion around the two-dimensional OA manifold in the absence of a frequency distribution. For multi-harmonic coupling, it similarly approximates the dynamical system on the infinite-dimensional Bernstein–Szegő (BS) manifold through a perturbative expansion around the finite-dimensional BS manifold in the absence of a frequency distribution. However, unlike the reduction to the OA manifold, the reduction to the BS manifold is itself approximate. Consequently, in the multi-harmonic case, an additional error arises from the approximation by the BS manifold. This theory enables a low-dimensional reduction with explicit error estimates of a kind not available in conventional rational-function approximations for populations of sinusoidally coupled phase oscillators with a Gaussian frequency distribution. It also enables a low-dimensional reduction for populations of phase oscillators with multi-harmonic coupling and a Gaussian frequency distribution. As a result, the present theory is expected to provide a powerful framework for investigating the effects of frequency distributions on populations of phase oscillators without relying on potentially unrealistic heavy-tailed distributions, such as those of rational-function form.

As a direction for future research building on this work, particularly in the case of sinusoidal coupling, one may consider replacing Cauchy noise with a parameter distribution of Cauchy form. The OA ansatz has been extended to systems with general parameter distributions \cite{Pietras2016}, and when the parameter distribution is Cauchy, the residue theorem is expected to reduce the system to a form similar to that of a system subject to Cauchy noise. It would be interesting to investigate whether the present method can also be applied to systems combining such a Cauchy parameter distribution with a Gaussian frequency distribution. Furthermore, for other frequency distributions whose moments of all orders are well defined, a decomposition analogous to the Fourier–Hermite decomposition used in this paper can be constructed using the corresponding family of orthonormal polynomials. Developing a similar low-dimensional reduction theory for more general frequency distributions would also constitute an important extension.

\section*{Acknowledgments}

The author thanks Hiroshi Kori for valuable discussions. This study was supported by the WINGS-FMSP program at the University of Tokyo.

\bibliographystyle{unsrtnat}
\bibliography{Ref_shortened}

\end{document}